\documentclass[12pt]{iopart}
\usepackage{iopams}
\usepackage{graphics}
\usepackage{graphicx}

\def\empile#1\above#2{\mathrel{\mathop{\kern 0pt#1}\limits_{#2}}}

\newcommand{\non}{\nonumber\\}

\newcommand{\sll}{\raise.15ex\hbox{$/$}\kern-.43em\hbox{$l$}}
\newcommand{\slepsilon}{\raise.15ex\hbox{$/$}\kern-.53em\hbox{$\epsilon$}}
\newcommand{\slvarepsilon}{\raise.15ex\hbox{$/$}\kern-.53em\hbox{$\varepsilon$}}\newcommand{\slL}{\raise.15ex\hbox{$/$}\kern-.53em\hbox{$L$}}
\newcommand{\slP}{\raise.15ex\hbox{$/$}\kern-.53em\hbox{$P$}}
\newcommand{\slp}{\raise.1ex\hbox{$/$}\kern-.63em\hbox{$p$}}
\newcommand{\slq}{\raise.1ex\hbox{$/$}\kern-.53em\hbox{$q$}}
\newcommand{\slv}{\raise.1ex\hbox{$/$}\kern-.63em\hbox{$v$}}
\newcommand{\slR}{\raise.15ex\hbox{$/$}\kern-.53em\hbox{$R$}}
\newcommand{\slQ}{\raise.15ex\hbox{$/$}\kern-.53em\hbox{$Q$}}
\newcommand{\slK}{\raise.15ex\hbox{$/$}\kern-.53em\hbox{$K$}}
\newcommand{\slk}{\raise.15ex\hbox{$/$}\kern-.53em\hbox{$k$}}
\newcommand{\slSigma}{\raise.15ex\hbox{$/$}\kern-.53em\hbox{$\Sigma$}}
\newcommand{\slcalP}{\raise.15ex\hbox{$/$}\kern-.63em\hbox{$\cal P$}}
\newcommand{\slA}{\raise.15ex\hbox{$/$}\kern-.73em\hbox{$A$}}
\newcommand{\slbfA}{\raise.15ex\hbox{$/$}\kern-.73em\hbox{${\imb A}$}}
\newcommand{\slpartial}{\raise.15ex\hbox{$/$}\kern-.53em\hbox{$\partial$}}
\newcommand{\sla}{\raise.15ex\hbox{$/$}\kern-.53em\hbox{$a$}}

\newcommand{\slb}{\raise.15ex\hbox{$/$}\kern-.53em\hbox{$b$}}
\newcommand{\slc}{\raise.15ex\hbox{$/$}\kern-.53em\hbox{$c$}}
\newcommand{\slD}{\raise.15ex\hbox{$/$}\kern-.53em\hbox{$D$}}
\newcommand{\slC}{\raise.15ex\hbox{$/$}\kern-.53em\hbox{$C$}}

\def\p{{\boldsymbol p}}
\def\q{{\boldsymbol q}}
\def\l{{\boldsymbol l}}
\def\k{{\boldsymbol k}}

\def\x{{\boldsymbol x}}

\def\wt{\widetilde}
\def\bs{\boldsymbol}

\begin{document}

\title{Heavy flavor production in pA collisions}

\author{H Fujii$^1$, F Gelis$^2$ and R Venugopalan$^3$}

\address{$^1$ Institute of Physics, University of Tokyo, Komaba, Tokyo 153-8902, Japan}
\address{$^2$ CEA/DSM/SPhT, Saclay, 91191, Gif-sur-Yvette Cedex, France}
\address{$^3$ Physics Department,  Brookhaven National Laboratory, Upton, NY 11973, USA}

\begin{abstract}
Heavy quark production in high-energy proton-nucleus (pA) collisions
is described in the framework of the Color Glass Condensate.
$k_\perp$ factorization is broken even at leading order albeit a more general factorization in pA holds at this order 
in terms of 2, 3 and 4 point correlators of Wilson lines in the nuclear target. The $x$-evolution of these correlators is 
computed in the large $A$ and large $N$ mean field limit of the Balitsky-Kovchegov equation. We show results for heavy quark 
production at  RHIC and LHC energies.
\end{abstract}

\section{Introduction}

Heavy flavor production in high energy processes is 
usually described in the collinear or $k_\perp$ factorization frameworks
of perturbative QCD.
Reactions with nuclear targets allow one to study 
new important features of 
multiple scatterings and parton saturation, which is 
characterized by the scale $Q_{sA}^2$\cite{IancuVrev}.
For a large nucleus at higher energy,
this scale can be large and comparable to the heavy quark mass $m_Q$,
so that 
$\Lambda^2_{\rm QCD} \ll m_Q^2 \lesssim Q_{sA}^2  \ll s $.
In this situation, realized at RHIC and definitely at future LHC,
we can investigate saturation physics through heavy quark
production with the generalized framework
of the Color Glass Condensate (CGC)
\cite{BlaizGV2,FujiiGV1,FujiiGVq1,KharzT1,KharzT4,AlbacK}.


In nucleus-nucleus collisions, heavy quarks are produced
by initial hard interactions and subsequently 
propagate in the hot flowing medium which modifies 
the $p_\perp$ spectrum of the quarks
and the quarkonium yield\cite{PnxElossC}.
In order to quantify these in-medium effects in hot matter,
proton/deuteron-nucleus (pA) collisions play a decisive role 
providing the baseline of the initial nuclear effects.
Here, we will compute heavy quark production in high energy pA
collisions using the CGC framework that 
incorporates both multiple scattering and small $x$ evolution effects.

\section{Quark pair cross section in the large N limit} 

In pA collisions, which is a prototype of a dilute-dense system,
the particle production cross-sections are known analytically
to the leading order in the strong coupling constant $ \alpha_s$
and the proton source $\rho_p$, 
but to all orders in the dense nuclear source
$g^2 \rho_A=O(1)$\cite{BlaizGV2}.
The cross-section of the quark pair production is written
in the CGC framework as 
\begin{eqnarray}
&&\frac{d \sigma_{q \bar{q}}}{d^2\p_\perp d^2\q_\perp dy_p dy_q}
=
\frac{\alpha_s^2 }{(2\pi)^6 C_F}
\non
&&\qquad\qquad\qquad\quad\times
\int\limits_{\k_{2\perp},\k_\perp}\!\!\!
\frac{\Xi(\k_{1\perp},\k_{2\perp},\k_{\perp})}
{\k_{1\perp}^2 \k_{2\perp}^2}
\;
\phi_{_A,y_2}^{q\bar{q},g}(\k_{2\perp},\k_\perp)
\;
\varphi_{p,y_1}(\k_{1\perp})
\; ,
\label{eq:cross-section-LN}
\end{eqnarray}
where $\k_{1\perp}=\p_\perp+\q_\perp - \k_{2\perp}$,
and $\k_{1,2\perp}$ and $y_{1,2}=\ln(1/x_{1,2})$ 
are the momenta and rapidities of the gluons coming
from the proton and the nucleus.
A shorthand notation
$\Xi(\k_{1\perp},\k_{2\perp},\k_{\perp})$ stands for the
matrix element squared, whose explicit form can be found
in Refs.~\cite{BlaizGV2,FujiiGV1,FujiiGVq1}.
In the leading twist approximation on both the proton and 
nuclear sides, this $\Xi$ reduces to the LO hard matrix element in the
$k_\perp$ factorization formalism\cite{GelisV1}.
The $\varphi_{p,y_1}(\k_{1\perp})$ 
is the $k_\perp$-dependent gluon distribution in the
proton, which is related to the usual gluon distribution
function via:
\begin{eqnarray}
{1 \over {4\pi^3} }
\int^{Q^2} d \k_{\perp}^2 \varphi_{p,y}(\k_{\perp})
\equiv
xG_p(x,Q^2)
\; .
\end{eqnarray}
The multi-gluon correlations in the large nucleus are encoded
in the expression (\ref{eq:cross-section-LN})
as the 3-point correlator (large $N$)
\begin{eqnarray}
\phi_{_A,_Y}^{q \bar{q},g}(\l_\perp,\k_\perp)
&\empile{=}\above{\rm LN}&
\pi R_A^2 \;
\frac{\l^2_\perp N}{4 \alpha_s} \;
S_Y(\k_\perp) \;
S_Y(\l_\perp-\k_\perp)  ,
\label{eq:3ptfn}
\end{eqnarray}
where $S_Y(\k_\perp)$ is the Fourier transform of 
the forward scattering amplitude of the right-moving dipole
$S_Y(\x_\perp)=
(1/N){\rm tr}\big<{\wt U}(\x_\perp){\wt U}^\dagger(0)\big>_Y $
with
\begin{eqnarray}
{\wt U}(\x_\perp)\equiv {\cal P}_+ \exp\left[-ig^2\int_{-\infty}^{+\infty}
dz^+ \frac{1}{{\bs\nabla}_{\perp}^2}\,\rho^a_{_A}(z^+,\x_\perp)\; t^a
\right] \; .
\end{eqnarray}
Here $t^a$ is the SU($N$) generator in the fundamental representation.

\begin{figure}[htb]
\begin{center}
\resizebox*{!}{2.8cm}{\includegraphics[angle=0]{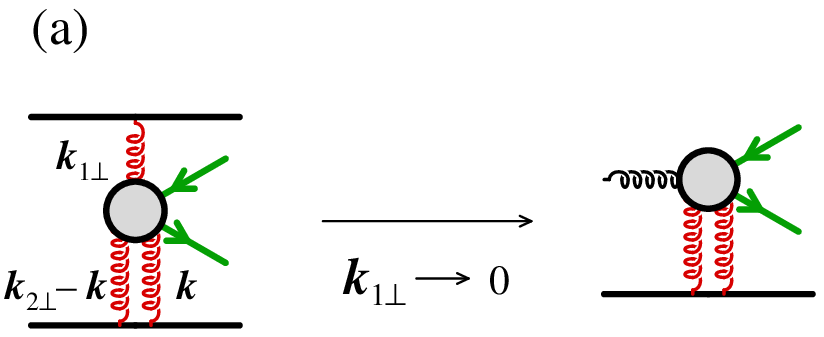}}
\qquad\qquad
\resizebox*{!}{4cm}{\includegraphics[angle=0]{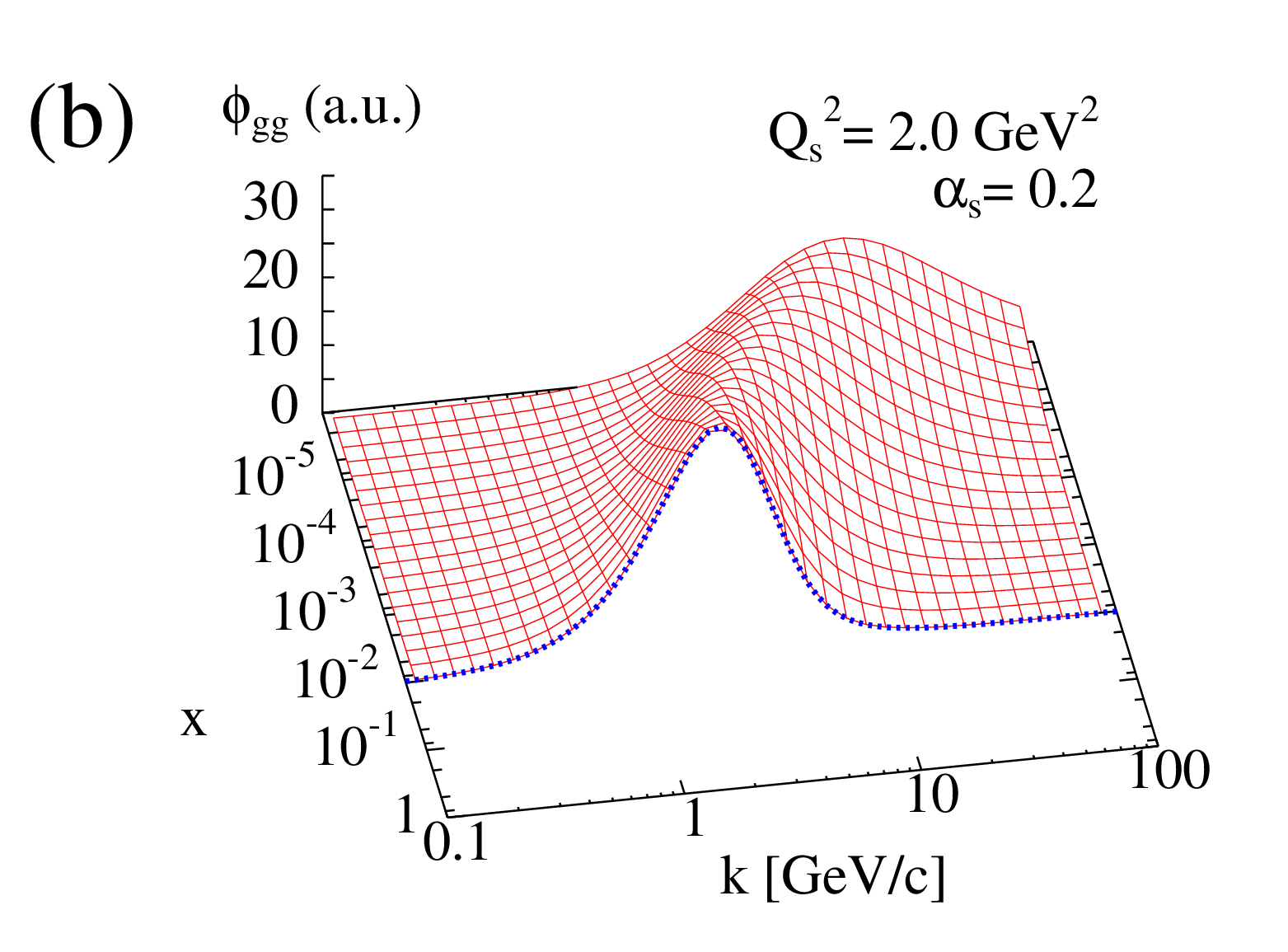}}
\end{center}
\caption{\label{fig:formalism}
(a) Collinear limit on the proton side. Blob stands for the
 interactions with the nucleus including multiple scatterings. (b)
Nuclear gluon distribution obtained from the BK equation with the MV
model initial condition.}
\end{figure}

On the proton side, the collinear approximation is more appropriate
because the typical $\k_\perp$ should be small --
$O(\Lambda_{\rm QCD })$ as long as $x_1$ is not too small,
especially in the forward rapidity (large $x_1$) region.
As noted in Refs.~\cite{GelisV1,BlaizGV2} the collinear limit on the
proton side is well-defined thanks to a Ward identity: one obtains 
(see Fig.~\ref{fig:formalism} (a))
\begin{eqnarray}
&&
\frac{d\sigma_{q\bar{q}}}{d^2\p_\perp d^2\q_\perp dy_p dy_q}
=
\frac{\alpha_s^2 }{4(2\pi)^4 C_F}
\non
&&\qquad\qquad\qquad\quad\times 
\int\limits_{\k_\perp}
\frac{\Xi^\prime(\k_{2\perp},\k_{\perp})}{\k_{2\perp}^2}
\;
\phi_{_A,y_2}^{q\bar{q},g}( \k_{2\perp},\k_\perp)
\;
 x_1 G_p(x_1,Q^2) ,
\label{eq:cross-section-LN-coll}
\end{eqnarray}
where $\k_{2\perp}=\p_\perp + \q_\perp$, and
\begin{eqnarray}
\Xi^\prime(\k_{2\perp},\k_{\perp})
\equiv
\lim_{|\k_{1\perp}|\to 0}
\int \frac{d\theta_1}{2\pi}
\frac{\Xi(\k_{1\perp},\k_{2\perp},\k_{\perp})}{\k_{1\perp}^2} \; .
\end{eqnarray}

In this work, we numerically solve
the Balitsky-Kovchegov (BK) equation 
for $S_Y(\k_\perp)$ \cite{Balitsky,Kovchegov}
with the McLerran-Venugopalan (MV)
model\cite{McLerrV1} initial condition at $x_0$=0.01,
and compute the 3-point correlator
$\phi_{A,y_2}^{q\bar{q},g}(\k_{2\perp},\k_\perp)$ 
(Fig.~\ref{fig:formalism} (b)).
This prescription is justified in the large $N$ and large $A$ limit.
For large-$x$ extrapolation, see Ref.~\cite{FujiiGVq1}.

\begin{figure}[hbt]
\begin{center}
\resizebox*{!}{4.7cm}{\includegraphics[angle=0]{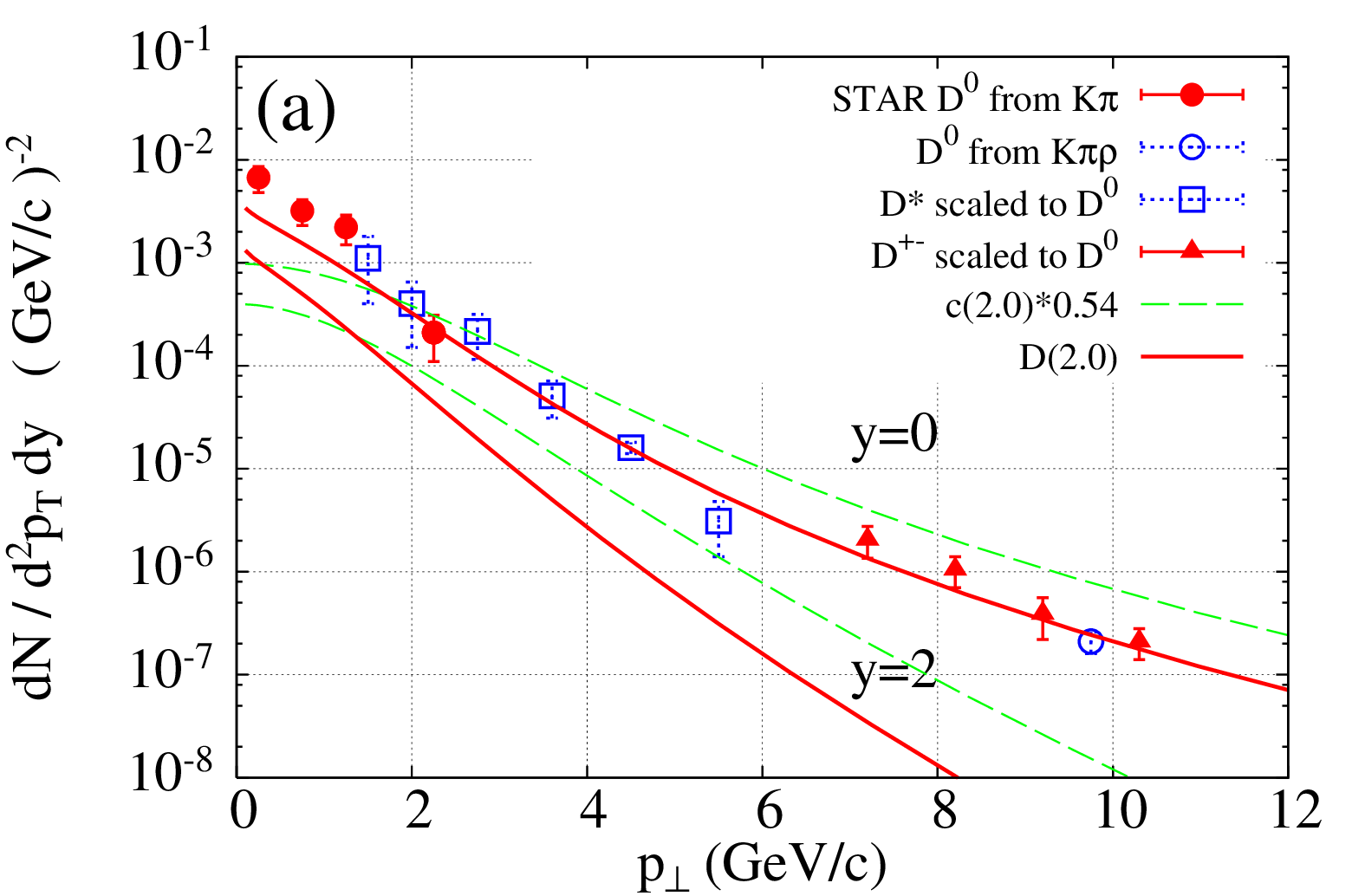}}
\qquad
\resizebox*{!}{4.5cm}{\includegraphics[angle=0]{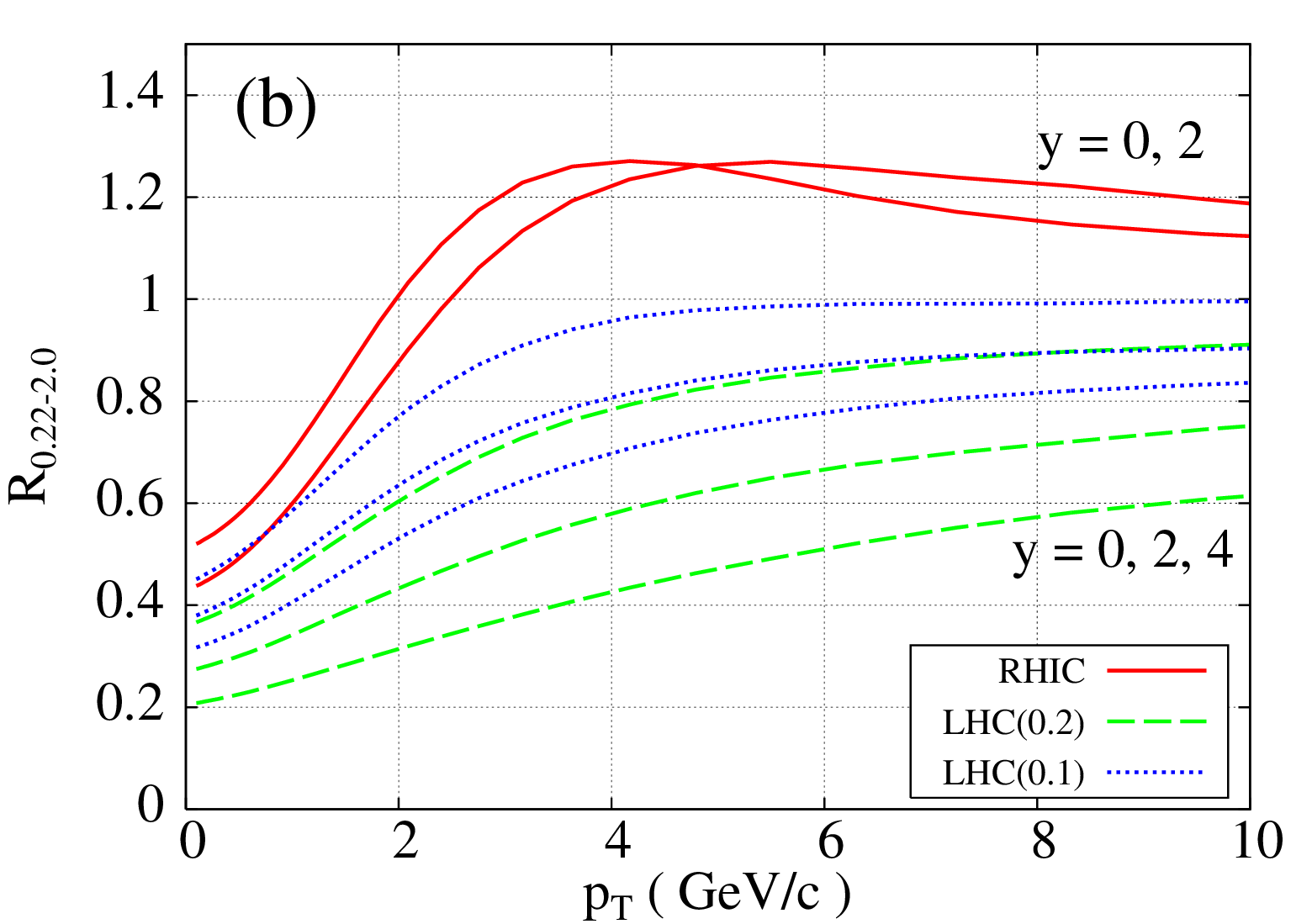}}
\end{center}
\caption{\label{fig:opencharm}
(a) Open charm spectrum at $\sqrt{s}=200$ GeV at $y$=0, 2 calculated
with $\alpha_s$=0.2 and $Q_{sA0}^2$=2GeV$^2$.
(b) Nuclear modification factor of $Q_{sA0}^2$=2 GeV$^2$ to 0.22
GeV$^2$ for D meson at $\sqrt{s}$=200 and
8500 GeV. In the latter case, results of the BK evolution with $\alpha_s$
=0.1 and 0.2 are compared.}
\end{figure}

\section{Open heavy flavor}

In Fig.~\ref{fig:opencharm} (a), we show
the single charm spectrum from eq.~(\ref{eq:cross-section-LN-coll})
with $Q_{sA}^2$=2 GeV$^2$ at $x=0.01$ and the BK evolution with fixed
$\alpha_s$=0.2. CTEQ6LO\cite{CTEQ6} is used for the proton.
The D$^0$ spectrum is obtained after convolution with Peterson's
fragmentation function $D(z)$ ($\epsilon=0.05$ and $z=p_{D\perp}/ p_{c\perp}$).
Preliminary data are taken from STAR~\cite{StarQM04,StarDAOpenC}.
We see that after the convolution
the D$^0$ spectrum is a little harder at lower $p_\perp$ than the data
while it fits in the higher $p_\perp$ region. 
Note that the b-quark contribution is not included in our calculation.
Fig.~\ref{fig:opencharm} (b) presents
the nuclear modification factor
of $Q_{sA}^2(x=0.01)$= 2 to 0.22 GeV$^2$,
at $\sqrt{s}$= 200 GeV ($y$=0, 2) and 8500 GeV ($y$=0, 2, 4).
We see at RHIC energy the Cronin peak moving to higher $p_\perp$ with
the rapidity $y$ because the charm pair production is dominated by the
moderate $x_2$ region.
Instead, at LHC energy we see larger suppression with $y$;
Note that fixed coupling $\alpha_s$=0.2 yields too large $\lambda$ of
$Q_s^2(x)\sim (1/x)^{\lambda}$ compared with HERA analysis.
The saturation effects can be seen even in the B meson productions at this energy.

\section{Quarkonium}

\begin{figure}[tbh]
\begin{center}
\resizebox*{!}{3.5cm}{\includegraphics[angle=0]{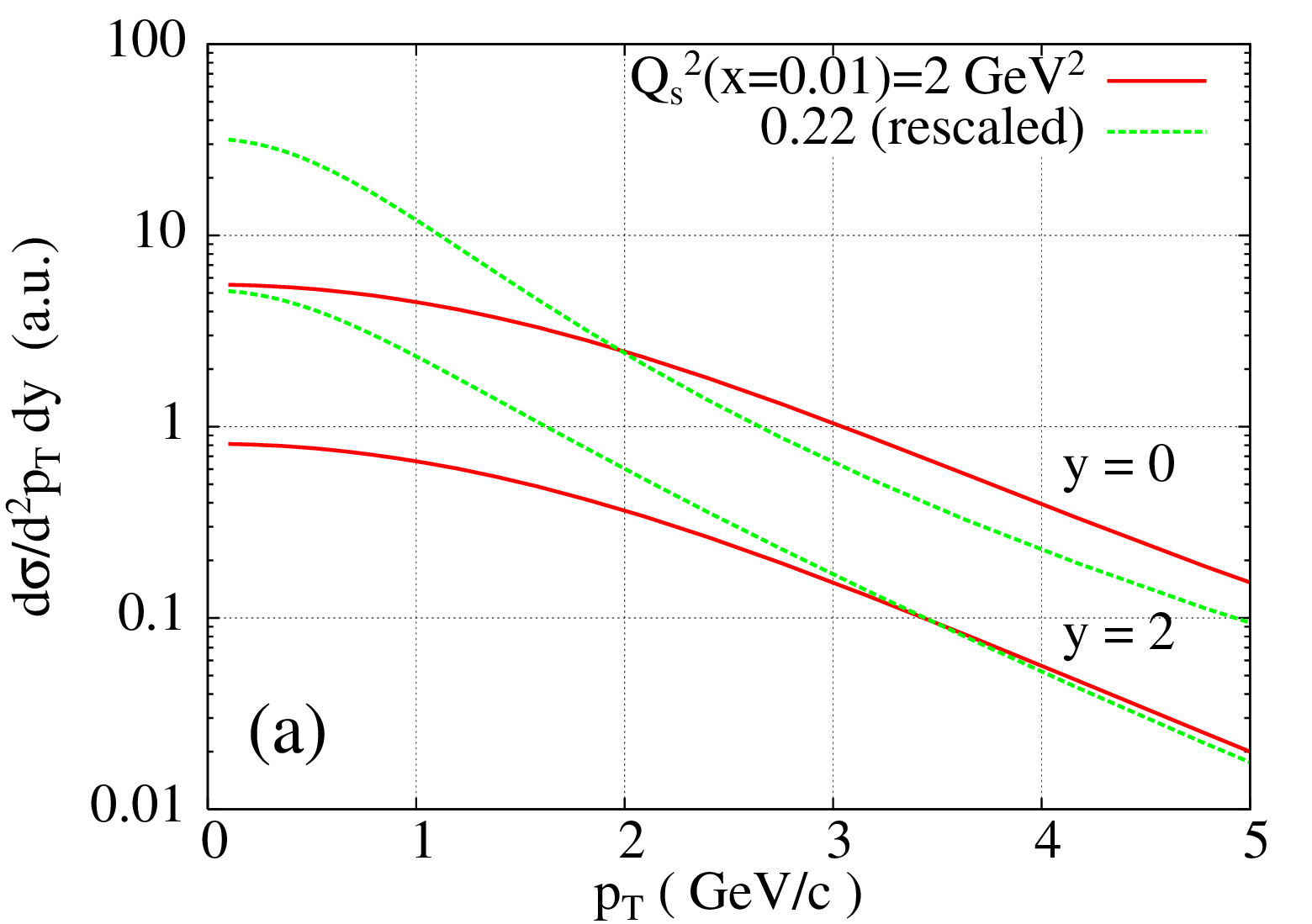}}
\resizebox*{!}{3.5cm}{\includegraphics[angle=0]{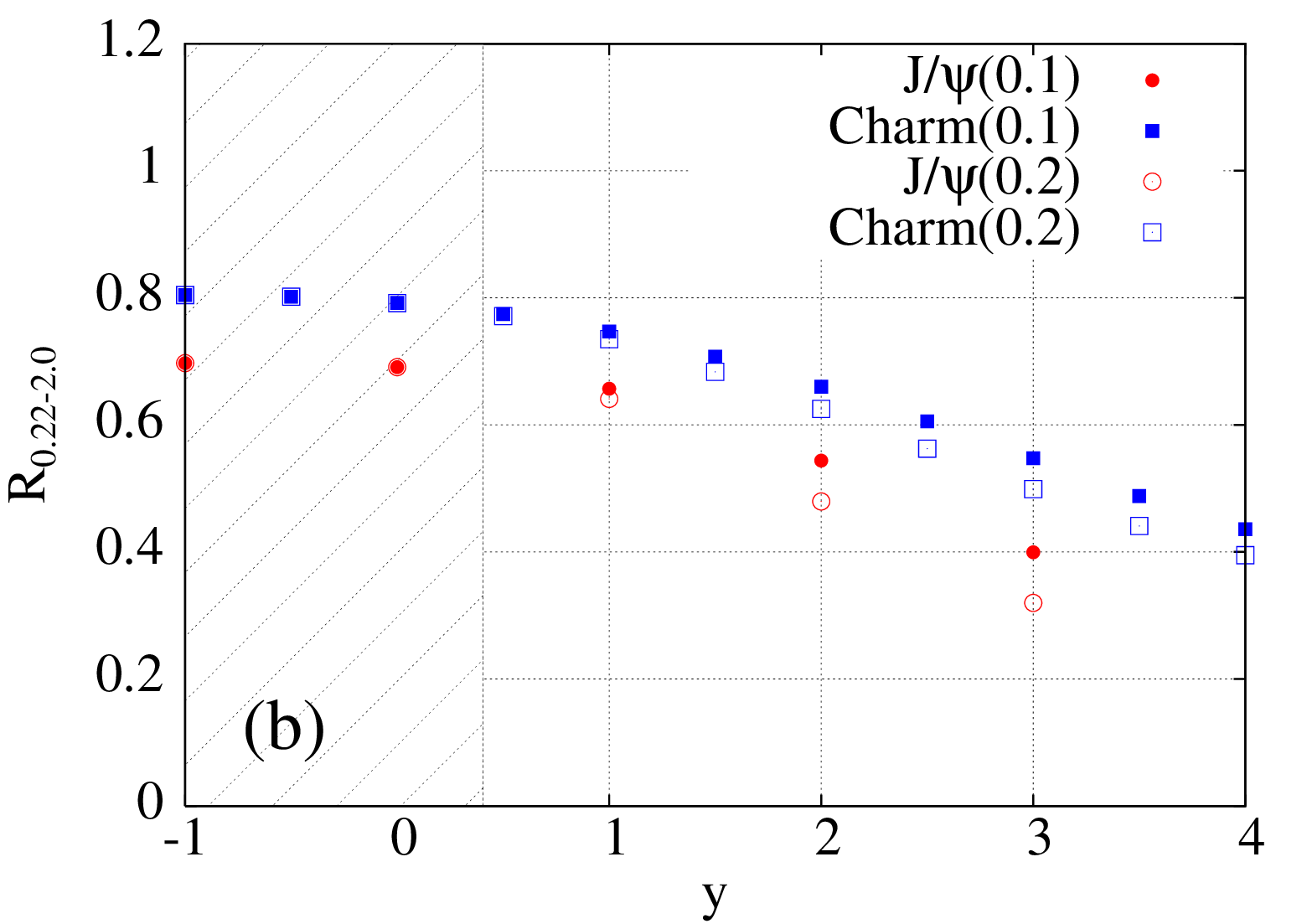}}
\resizebox*{!}{3.5cm}{\includegraphics[angle=0]{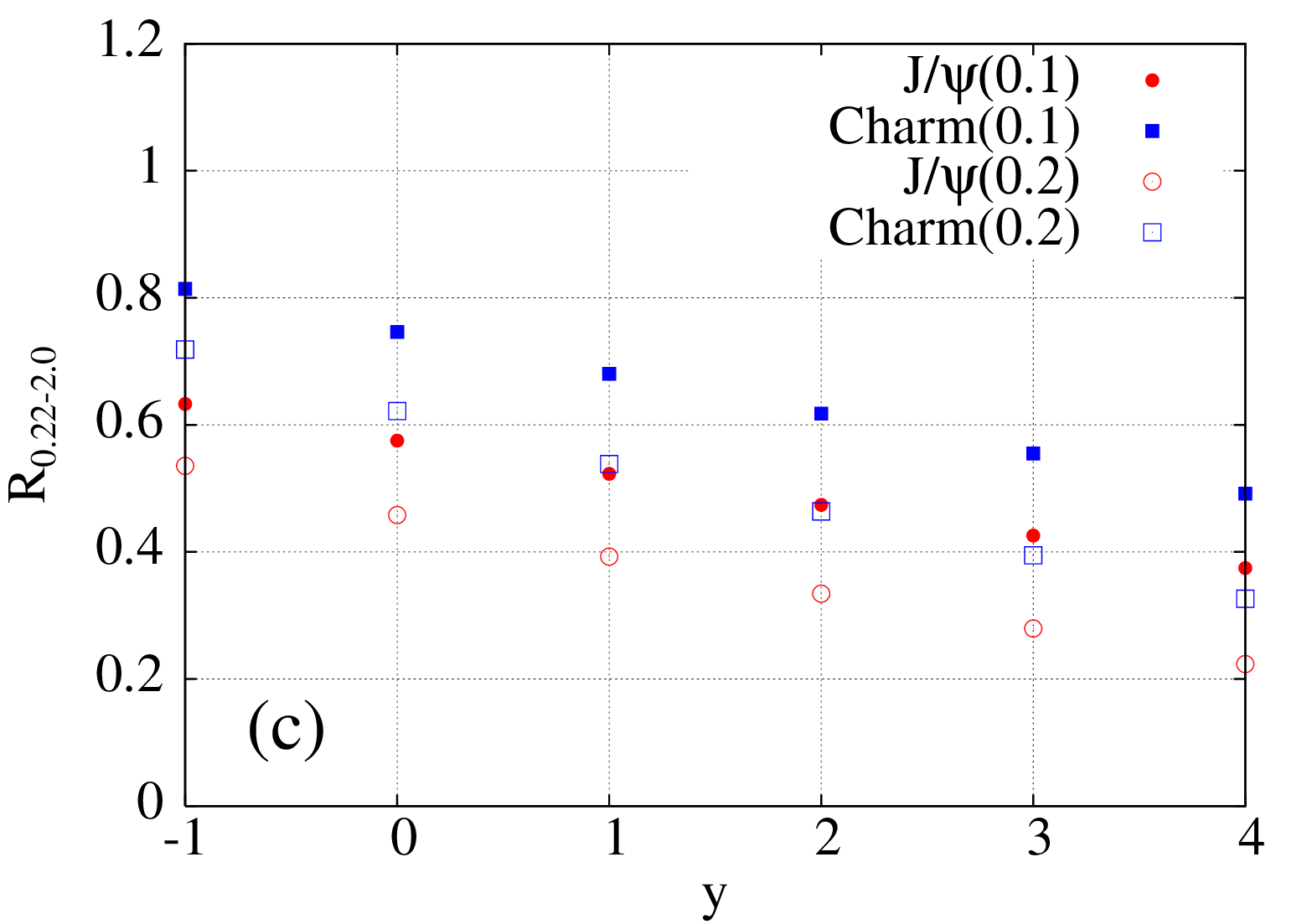}}
\end{center}
\caption{\label{fig:jpsi}
(a) J/$\psi$ spectrum for $Q_{sA0}^2$=2 and 0.22 GeV$^2$ at $\sqrt{s}$= 200 GeV.
(b) Nuclear modification factor of J/$\psi$ and charm quark ($\alpha_s$=0.1 and 0.2).
Hatched region correspoinds to $x_2 \gtrsim 0.01$.
(c) The same as (b) at $\sqrt{s}$= 8500 GeV.}
\end{figure}

The J/$\psi$ production cross-section is computed for $Q_{sA0}^2$=2
and 0.22 GeV$^2$ in the color evaporation model in
Fig.~\ref{fig:jpsi},
where we assume that the bound state formation takes place outside of
the nucleus, and is thus not modified compared to pp.
The J/$\psi$ absorption in the target is not necessary.
The cross-section is supressed (enhanced) in the small (large)
$p_\perp$ region at RHIC energy due to the multiple scattering of the pair
(Fig.~\ref{fig:jpsi} (a))\cite{FujiiM1}.
At forward rapidities $dN/dy$
is more suppressed due to the multiple
scattering {\it and} saturation, which may be compared with
the data\cite{PnxDAJpsi}. 
Note that in the hatched
region $y \lesssim 0.5$ in Fig.~\ref{fig:jpsi} (b), 
$dN/dy$ is largely determined by the larger $x_2$ part of $\phi_{A,y_2}^{q\bar{q},g}$,
which is extrapolated from the MV model.
In Fig.~\ref{fig:jpsi} (c), we show the suppressions of $dN/dy$ for the
J/$\psi$ and the charm quark at LHC energy. 
We expect that the difference between them 
represents the effect of the multiple scattering on the bound state formation.
It is interesting to note that this difference depends on the rapidity
$y$ only weakly.

\end{document}